\documentclass[12pt]{iopart}
\usepackage{amsfonts}
\usepackage{iopams}
\usepackage{graphicx}
\usepackage{bm}
\usepackage{color}
\usepackage{amstext}
\usepackage{amssymb}
\usepackage{mathrsfs}
\usepackage{grffile}
\usepackage[label font={bf, Large}]{subfig}
\usepackage{floatrow}

\begin{document}

\title{Learning What a Machine Learns in a Many-Body Localization Transition}
\author{Rubah Kausar$^1$,Wen-Jia Rao$^{2,*}$, and Xin Wan$^{1,3}$}
\address{$^1$Zhejiang Institute of Modern Physics, Zhejiang University, Hangzhou 310027, China \\ $^2$School of Science, Hangzhou Dianzi University, Hangzhou
310027, China \\$^3$CAS Center for Excellence in Topological Quantum Computation, University of Chinese Academy of Sciences, Beijing 100190, China\\ $^*$ E-mail: wjrao@hdu.edu.cn}
\begin{abstract}
We employ a convolutional neural network to explore the distinct phases in
random spin systems with the aim to understand the specific features that
the neural network chooses to identify the phases. With the energy spectrum
normalized to the bandwidth as the input data, we demonstrate that a network
of the smallest nontrivial kernel width selects level spacing as the
signature to distinguish the many-body localized phase from the thermal
phase. We also study the performance of the neural network with an increased
kernel width, based on which we find an alternative diagnostic to detect
phases from the raw energy spectrum of such a disordered interacting system.
\end{abstract}


\maketitle


\section{Introduction}

Recent research has established the existence of two generic phases in
isolated quantum many-body systems: the thermal phase and many-body
localized (MBL) phase~\cite{Gornyi2005,Basko2006}. Ergodicity is preserved
in the thermal phase, while in the MBL phase localization persists in the
presence of weak interactions. The difference between the thermal and MBL
phases exhibits in many aspects, such as quantum entanglement. A thermal
system can act as the heat bath for its own subsystem, hence the entanglement is
extensive and satisfies a volume law. An MBL system, however, yields small
entanglement that scales with the area of subsystem boundary. More recently,
much attention is drawn to the study of entanglement spectrum (ES)~\cite%
{li-haldane}, which is the eigenvalue spectrum of the reduced density matrix
of a subsystem. ES contains more information than its von Neumann entropy --
the entanglement entropy -- which is a single number. The variance of the
entanglement entropy and its evolution after a local quench from an
exact eigenstate, together with the spectral statistics of ES, are all
promising tools in the study of the MBL phase~\cite%
{Bardarson2012,Kjall2014,Geraedts2016,Geraedts2017,Yu2016,Yang2015,Serbyn16,Gray2017,Li2015,Maksym2015}%
.

Modern developments in machine learning (ML)~\cite{Goodfellow} provides a
new paradigm to study phases and phase transitions in condensed matter
physics. In computer science, ML is an efficient algorithm to extract hidden
features in data, such as figures, to make predictions about the nature of
new ones. This is similar to the study of phase transitions, where we use
(local or non-local) order parameters to distinguish different phases. ML
includes both unsupervised and supervised methods. Unsupervised learning is
a collection of exploratory methods that extract the hidden patterns in the
input data without prior knowledge of desired output. Whereas in supervised
learning the input data are accompanied by matching labels, and a machine is
trained to recognize patterns and predict correct labels. A significant
amount of works have been devoted in using ML methods to study equilibrium
phase transitions~\cite%
{Tanaka17,Carrasquilla17,Torlai16,YiZhang16,YiZhang17,Liu17,Wang16,Chng17,Morningstar17,Ponte17,Li17,Ohtsuki16,Hu17,ZYLi,Rao18,CNN-archi,Khatami}%
.

For the MBL physics, ML has been successfully employed to study the MBL
transition point, mobility edge, and the evolution of initial state~\cite%
{Nieuwenburg17,Nieuwenburg172,Venderley17,Schindler17,Hsu18,Doggen}.
In these works, ES is a popular choice of the input
data~\cite{Nieuwenburg17,Venderley17,Schindler17,Hsu18},
The choice, as was explicitly argued in Ref.~\cite{Hsu18},
builds on past studies that established ES as a sensitive probe
of the phases in the MBL systems, while ML successfully
extracts relevant features from the complex pattern of ES.
Unfortunately, such a common practice, unlike conventional studies in physics,
does not reveal what are the relevant features in the spectral pattern.
In addition, ES is obtained by pre-processing wave functions and is, therefore,
high-level data, while the general success of ML hints that we should be
able to use \textquotedblleft lower level\textquotedblright physical quantity,
such as wave function or energy spectrum.

Given the energy spectrum of a many-body system, the most widely-used
statistical quantity, i.e. a relevant feature, is the distribution of level spacings (gaps between
nearest levels). Eigenstates in the thermal phase are extended with finite
overlap with each other, resulting a correlated energy spectrum; while the
levels are independent in MBL phase. Consequently, as predicted by random
matrix theory~\cite{Haake2001} the nearest level spacings will follow a
Wigner-Dyson (WD)~\cite{Wigner,Dyson,Dyson1962} distribution in the thermal
phase, while a Poisson distribution is expected in MBL phase. This
difference holds the key to understand the transition from non-integrable to
integrable systems and quantum chaos and is widely used in studying the MBL
transition~\cite{Pal2010,Johri2015,Luitz2015,Moore2016,Regnault2016,C2}.
Practically, when counting the level spacings, we have to make the density
of states (DOS) uniform, which is commonly achieved by unfolding the
spectrum or by picking the middle part of spectrum where the DOS is almost
uniform. Though, ambiguity may arise in the unfolding strategy~\cite%
{Gomez2002}.

The neural network based ML algorithm, on the other hand, allows a machine
to learn MBL transition directly from unprocessed energy spectra. One of us
has shown that a simple three-layer feed-forward neural network can
correctly capture the MBL transition point in random spin systems, with raw
energy spectrum being the training data~\cite{Rao182}. Mathematically, the
fully-connected neural network provides a complicated nonlinear operation on
the energy levels that contains a large number of parameters, which prevents
one from peeking into the ML process and taking advantage of the ML
knowledge in future studies. Such a criticism is often heard among ML
skeptics, whose doubts are fully justified because modern deep neural
networks are designed to recognize patterns, rather than to understand
physics related to the patterns.

In this study we digress from the orthodox ML objectives to explore whether
we can understand \textquotedblleft what\textquotedblright\ a machine learns
via a deep neural network. In other words, we are not satisfied with the
neural network being an oracle that predicts the outcome; we want to know
how and based on what the oracle predicts. For this purpose, we employ a
convolutional neural network (CNN) to study the MBL transition in random
spin systems. Specific to the MBL transition, we ask what design is needed
for a neural network to develop the idea to distinguish the thermal and MBL
phases by nearest-neighbour level spacing. By increasing the complexity of
the network, we explore what can be an improved indicator for the
distinction from the neural network point of view.


\section{Models and Methods}

The canonical model to study the MBL phenomena in one dimension (1D) is the
spin 1/2 Heisenberg chain with random external fields~\cite{MBL_intro},
whose Hamiltonian is given by
\begin{equation}
\mathrm{H}=\mathrm{J}\sum_{i}\mathbf{S}_{i}\boldsymbol{\cdot }\mathbf{S}%
_{i+1}+\sum_{\alpha =x,z}\sum_{i}\varepsilon _{i}^{\alpha }S_{i}^{\alpha }.
\label{Eq:rfhm_Model}
\end{equation}%
Here, $S$ is the spin-1/2 operator at each site. The isotropic interaction $%
J $ couples nearest neighbouring spins. The disorder is introduced via a
random field that couples to the $x$ and $z$ component of the spin operator.
Such a random field is modelled by making $\varepsilon _{i}^{\alpha }$
random and $\varepsilon _{i}^{\alpha }\in \left[ -h,h\right] $ is sampled
from a uniform distribution of width $2h$. In this work we set the
interaction strength $J$ to be unity and implemented periodic boundary
conditions.

The random matrix theory (RMT) pioneered by Wigner and Dyson \cite{Haake2001}
in $1960$s to understand the behaviour of complex nuclei established a deep
connection between the symmetries of the Hamiltonian and the statistical
properties of the eigenvalue spectrum. For instance, the system with time
reversal invariance is represented by a Hamiltonian matrix that is symmetric
and real, which is invariant under orthogonal transformation, hence belongs
to the Gaussian orthogonal ensemble (GOE). Note that our model Eq.(\ref%
{Eq:rfhm_Model}) breaks time-reversal symmetry due to the external field,
while there remains an anti-unitary symmetry comprised of time reversal and
a rotation by $\pi $ of all spins about $z$ axis which leave the Hamiltonian
unchanged, hence belonging to GOE. The Hamiltonian with spin rotational
invariance while breaks time reversal symmetry is represented by matrix that
belongs to the Gaussian unitary ensemble (GUE), while Gaussian symplectic
ensemble (GSE) represents systems with time reversal symmetry present but
broken spin rotational symmetry. All these ensembles describe thermal phase
that has finite correlations between different energy levels, and there
exist characteristic features that are only determined by the symmetry while
independent of the microscopic details.

Among various features of RMT, the mostly used one is the distribution of
nearest level spacings $\mathcal{P}(s)$, where $s$ is the normalized spacing
$E_{i+1}-E_{i}$ between nearest levels. For our model Eq.~(\ref%
{Eq:rfhm_Model}), it can be proven that in the thermal phase with small
disorder, the nearest level spacings follows a GOE distribution $\mathcal{P}%
( s) =\frac{\pi s}{2}\exp \left( -\frac{\pi s^{2}}{4}\right) $, reflecting
the repulsion between levels. On the other hand, in a fully localized phase
with large disorder, all the energy levels become independent, the nearest
level spacings distribution evolves into the Poisson distribution $\mathcal{P%
}(s)=\exp \left( -s\right) $. The level spacing has been proved as a
powerful tool to explore the behaviour of complex systems such as disordered
systems~\cite{MIT,single-particle,MB_Schklovskii,Avishai2002}, chaotic \cite%
{chaotic} and quasi periodic systems \cite{quasi1}.

\begin{figure}[tbp]
\centering
\sidesubfloat[]{\includegraphics[width=0.4\columnwidth]{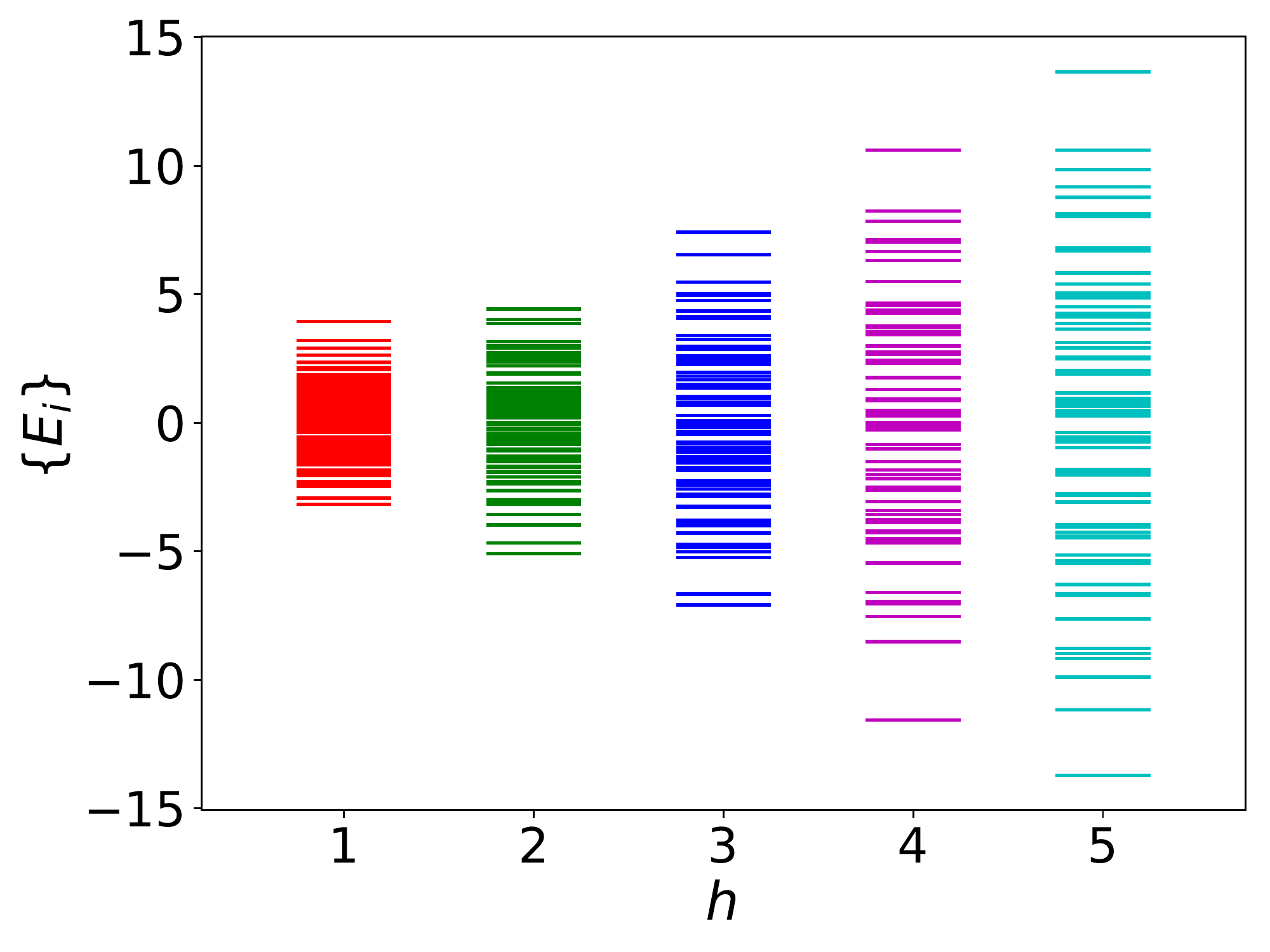}} %
\sidesubfloat[]{\includegraphics[width=0.4\columnwidth]{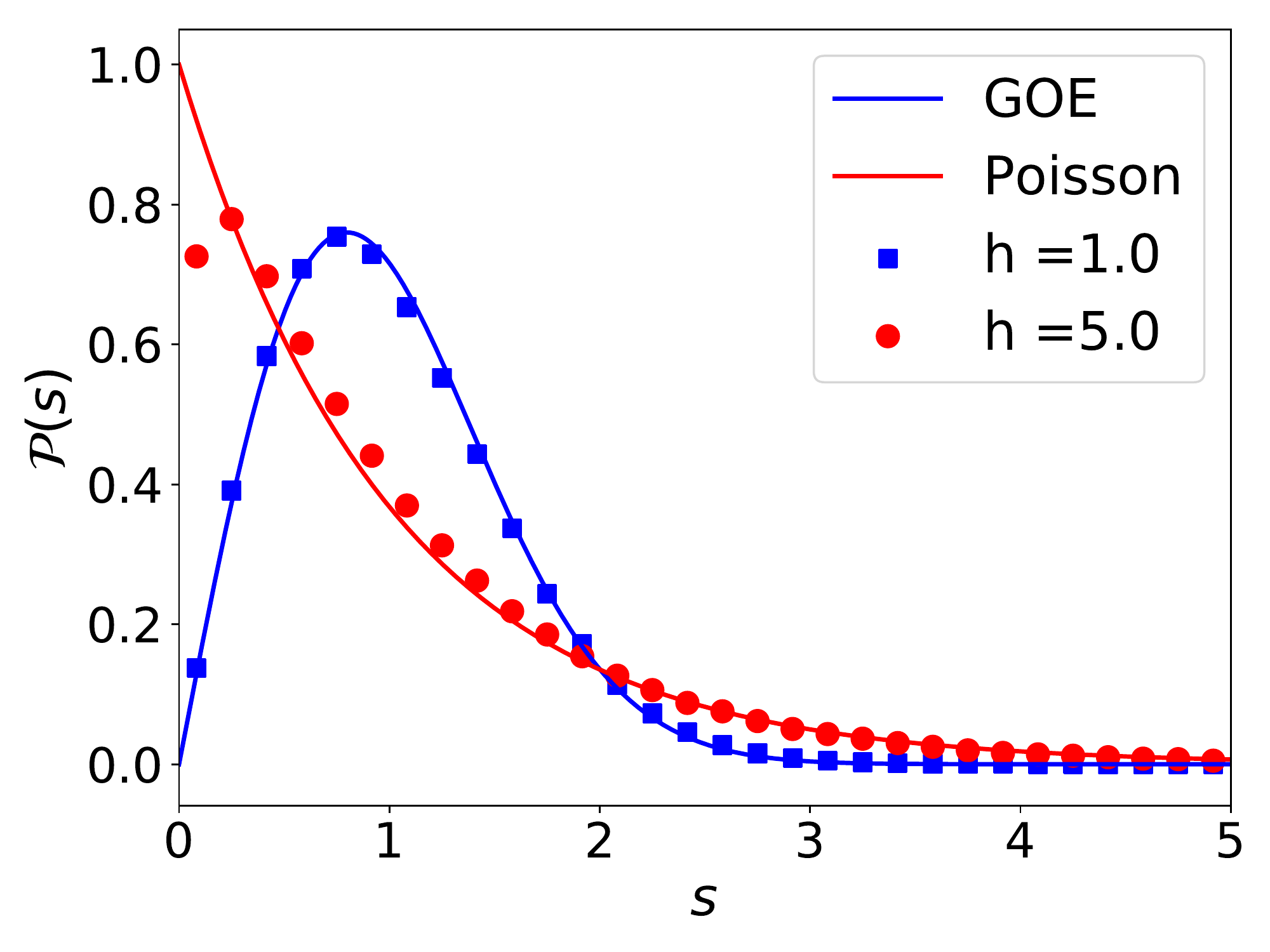}}
\caption{(a) Energy spectrum of random field Heisenberg model for $L=6$ at
different disorder strengths. (b) Comparison of the level spacing
distribution $\mathcal{P}(s)$ for the random field Heisenberg model. At
small disorder ($h=1$), $\mathcal{P}(s)$ follows the GOE distribution, while
at larger disorder ($h=5$) it has a Poisson distribution.}
\label{Fig:A}
\end{figure}

We plot representative energy spectra of the Hamiltonian in Fig.~\ref{Fig:A}%
(a), whose bandwidth increases with disorder strength $h$. More importantly,
in all cases, the levels are denser in the middle part of the spectrum,
hence the density of state (DOS) is more uniform. For this reason we choose
middle part of the spectrum to do level statistics. The evolution of the
level spacing distribution at low ($h=1$) and high ($h=5$) disorder strength
is shown in Fig.~\ref{Fig:A}(b), which is obtained after implementing the
unfolding procedure \cite{Kudo2004,unfold1,unfold2}. We clearly see that at
large disorder strength ($h=5$) the level spacing distribution is Poissonian
and at small value of disorder ($h=1$) the distribution follows GOE. The
fitting for $h=5$ has noticeable deviations around $s\rightarrow 0$ in the
finite system, where there remains small but finite correlation between
nearby eigenstates. Past studies estimate a critical $h_{c}\approx 3$~\cite%
{Nieuwenburg17,Regnault2016,C2} for the transition from the thermal phase to
the MBL phase.

Recent studies show that even though a machine has no knowledge of level
spacing in random matrix theory, a fully-connected feed-forward neural
network can, nevertheless, detect the MBL transition in random spin systems,
with the raw energy spectra in small systems as training data. However, the
large number of parameters in this network makes it difficult to extract
\textquotedblleft what\textquotedblright\ machine learns, hence, in this
work, we employ the convolutional neural network (CNN) \cite{CNN-archi} to
study the MBL transition.

\begin{figure}[tbp]
\centering
\sidesubfloat[]{\includegraphics[width=0.35\columnwidth]{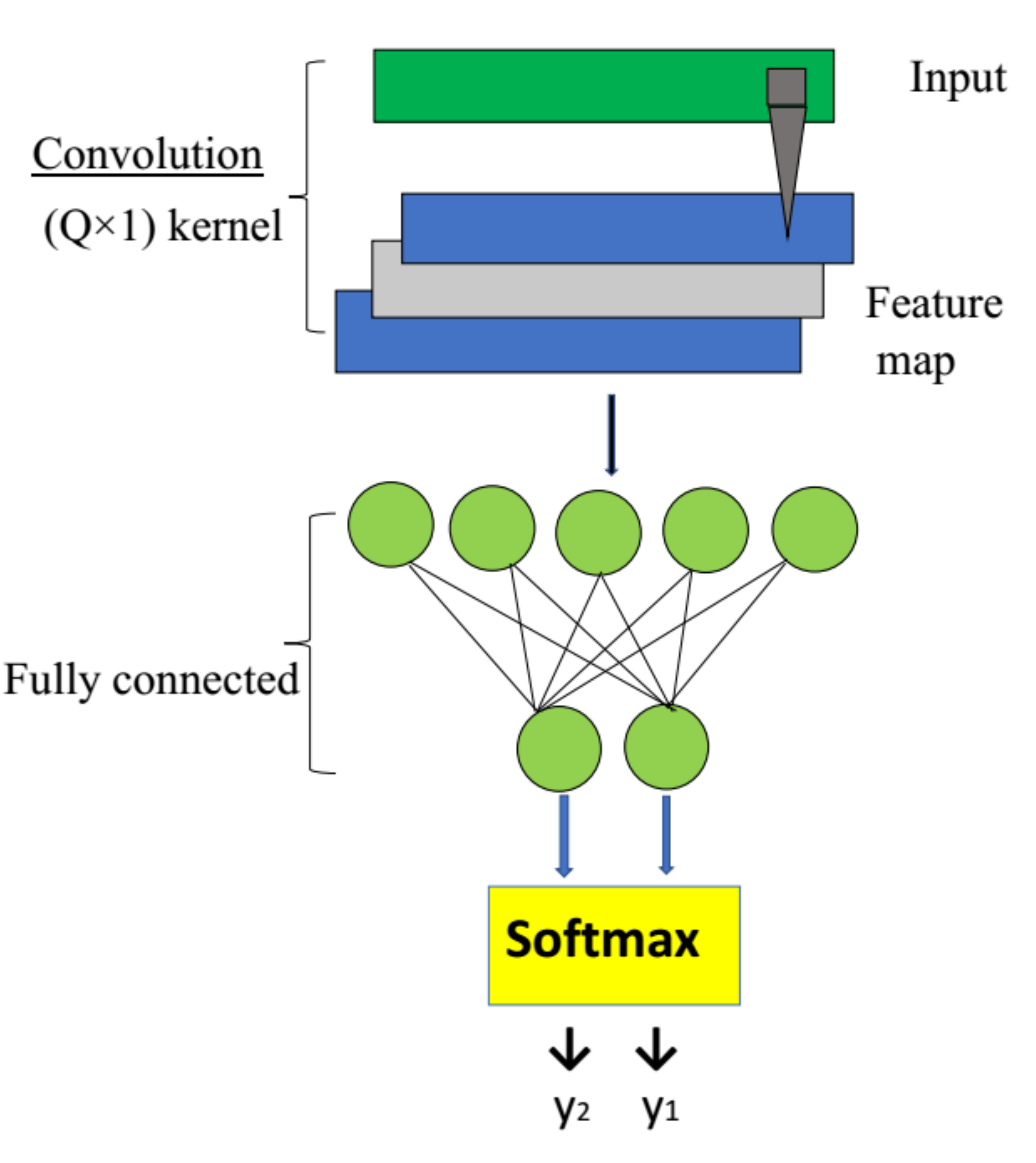}} %
\sidesubfloat[]{\includegraphics[width=0.45\columnwidth]{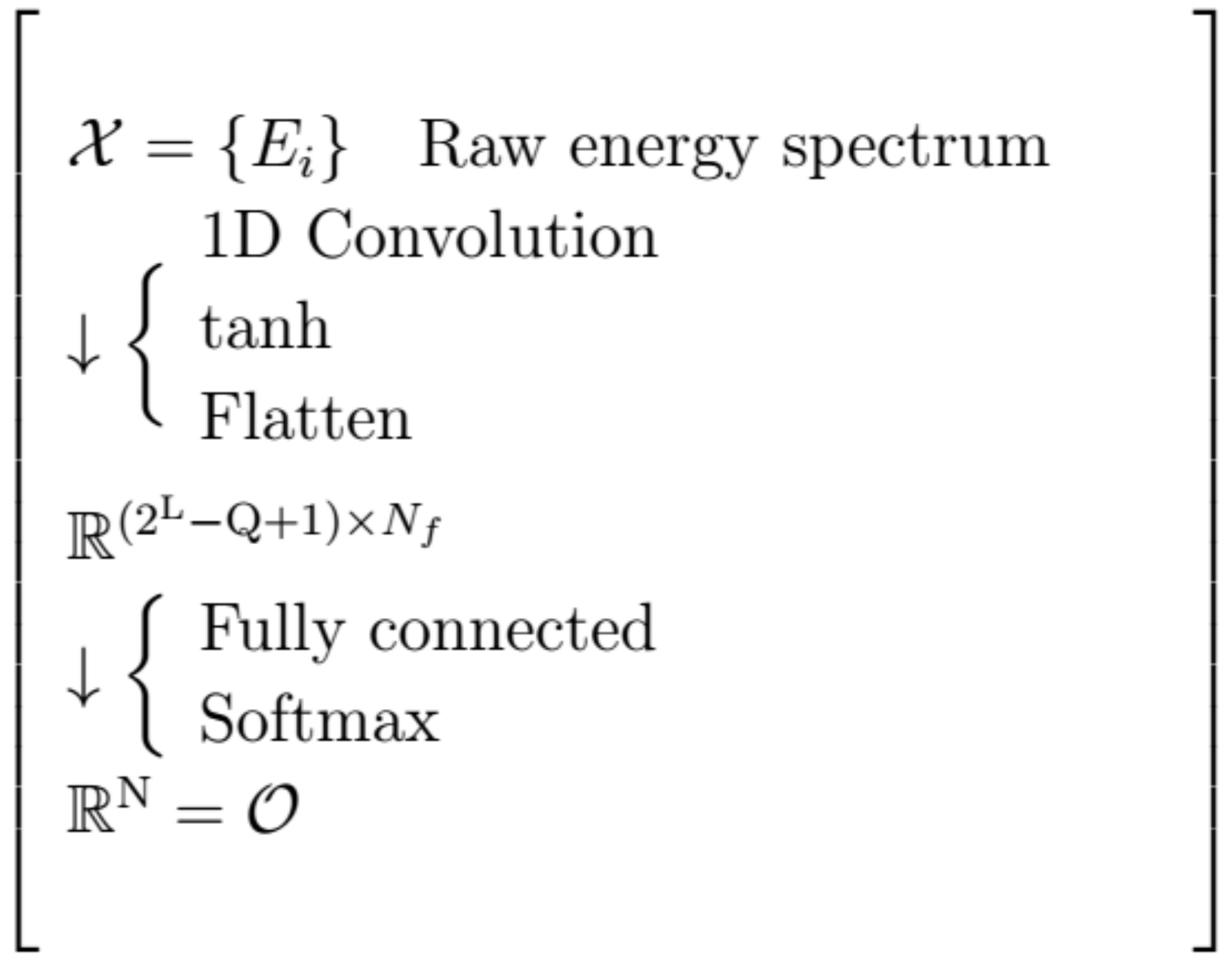}}
\caption{(a) Sketch of the CNN architecture used in the present work. (b)
The flow of data in the CNN model.}
\label{Fig:B}
\end{figure}

The CNN model $\mathcal{M}_{\mathrm{CNN}}$ consists of a convolution layer
and a fully connected layer as shown Fig.~\ref{Fig:B}(a). In the convolution
layer the input data $\mathcal{X}$ is convolved with $N_f$ filters $\mathcal{%
F}^{\beta}$ (also referred as kernel), where $\beta = 1, 2, \dots, N_{f}$.
Mathematically, the 1D convolutions performed by our CNN model can be
expressed as,
\begin{equation}
\mathrm{Z}^{\beta }=(\mathcal{X}\ast \mathcal{F}^{\beta })(k)=\sum_{i=1}^{Q}%
\mathcal{X}(k+i)\mathcal{F}^{\beta }(i),\quad  \label{Eq:Stride}
\end{equation}%
where $\mathrm{Z}^{\beta }$ is the $\beta$th feature map obtained as the
result of the convolution process and $Q$ is the width of the kernel. Filter
weights constitute a set of $N_{f}\times Q$ parameters in the convolution
layer, which are optimized during the training process. These resulting
feature maps are processed by a nonlinear activation function, whose output
is flattened and passed to the fully connected layer without pooling. By a
linear map with weights $\mathcal{W}$ the flattened output yields $\tilde{y}%
_{1}$ and $\tilde{y}_{2}$, which correspond to the two phases. The set of
weights are also optimized in the training process. Finally, the probability
of being in either of the phases is obtained by a softmax function
\begin{eqnarray}
y_{1}^{\mathrm{CNN}} &=& \frac{\exp(\tilde{y}_{1})}{\exp(\tilde{y}_{1})+\exp(%
\tilde{y}_{2})} \\
y_{2}^{\mathrm{CNN}} &=& \frac{\exp(\tilde{y}_{2})}{\exp(\tilde{y}_{1})+\exp(%
\tilde{y}_{2})}
\end{eqnarray}
The flow of the data in the CNN model is summarized in Fig.~\ref{Fig:B}(b).

To classify the thermal and MBL phases, we train the neural network in a
supervised learning scheme with a collection of raw eigenvalue spectra
obtained via diagonalizing Eq.~(\ref{Eq:rfhm_Model}). In other words, we
label the spectra with the corresponding phases and train the network to
extract relevant features from the input data. In particular, the CNN is
trained with the mini-batch gradient descent method. The optimization
algorithm searches for an optimal set of parameters that minimizes the cross
entropy
\begin{equation}
\mathcal{E}=-\sum_{I=1}^{N_{b}}\sum_{i=1}^{2}y_{I,i}\log y_{I,i}^{\mathrm{CNN%
}},
\end{equation}%
where $y_{i}=0$ or $1$ is the true phase label of the $I$th sample, and $%
N_{b}$ is the size of batch during one training. After the training, the
neural network can use its gained knowledge to predict or validate the class
for a new set of data. The performance of the network depends on the model
of the network, as well as the quality of training. It should be pointed out
that, the number of parameters in a CNN is large (although significantly smaller
than that of fully-connected NN), hence finding a global minimum of the cross
entropy is almost impossible. Nevertheless, our aim is not the precise values of
the optimized parameters or the best performance, but a numerical trend in the
parameters that enables the machine to extract non-trivial physics.


\section{CNN Training Results}

We begin by understanding what a CNN learns to distinguish phases,
explicitly, the thermal phase at low disorder and the MBL phase at high
disorder from energy spectra, as displayed in Fig.~\ref{Fig:A}(a), in the random spin
system. We assume no prior knowledge of the exact transition point and
numerically generate the raw energy spectrum $\left\{ E_{i}\right\} $ of the
Heisenberg model deep in each phase. Explicitly, we collect data for the
thermal phase (labelled as 0) in the range $1.0 \leq h \leq 1.4$, and for
the MBL phase (labelled as 1) in the range $4.6 \leq h \leq 5.0$. In each
region we generate $1000$ samples of $\left\{ E_{i}\right\} $ with parameter
interval $\Delta h = 0.1$. The conventional wisdom is that the energy
spectra in the two phases can be distinguished by nearest-neighbour level
spacing, as we demonstrate in Fig.~\ref{Fig:A}(b). In the CNN training, we
assume we have no knowledge of the level statistics and feed all the
labelled data to the CNN for a supervised learning. We then analyze the
kernel parameters to gain knowledge on how the neural network filter the
energy spectra to distinguish the two phases.

\subsection{Filters whose kernel width is 2}

An energy spectrum is a 1D set of data, so we use 1D filters with kernel
width $Q$. The simplest nontrivial case is $Q = 2$. Therefore, we start our
training with the CNN architecture that has $2\times 1$ filters, i.e. 1D
filters whose kernel width is 2. In the convolution operation the CNN
filters extract features from nearest neighbouring levels.

The feature map $\mathrm{Z}^{\beta }$ generated by convolution with a filter
$\mathcal{F}^{\beta } = ( \mathcal{F}_{1}^{\beta }, \mathcal{F}_{2}^{\beta
}) $ on two neighbouring eigenvalues $E_{i}$ and $E_{i+1}$ is,
\begin{equation}
\mathrm{Z}_{i}^{\beta }=\mathcal{F}_{1}^{\beta }E_{i}+\mathcal{F}_{2}^{\beta
}E_{i+1},  \label{Eq:conv}
\end{equation}%
where $\beta$ indicates the channel index. For each collection of data we
run the training process $\mathcal{T}_{\mathscr{L}}$ $1000$ times. During
each $\mathcal{T}_{\mathscr{L}}$, the network parameters are initialized
stochastically from a truncated normal distribution having mean $\mu =0$ and
standard deviation $\sigma =0.1$. All testing accuracies are close to $100\%$%
, suggesting that distinguishing the thermal phase from the MBL phase is an
easy task for even the simplest CNN architecture.

\begin{figure}[tbp]
\centering		
\sidesubfloat[]{\includegraphics[width=0.4\columnwidth]{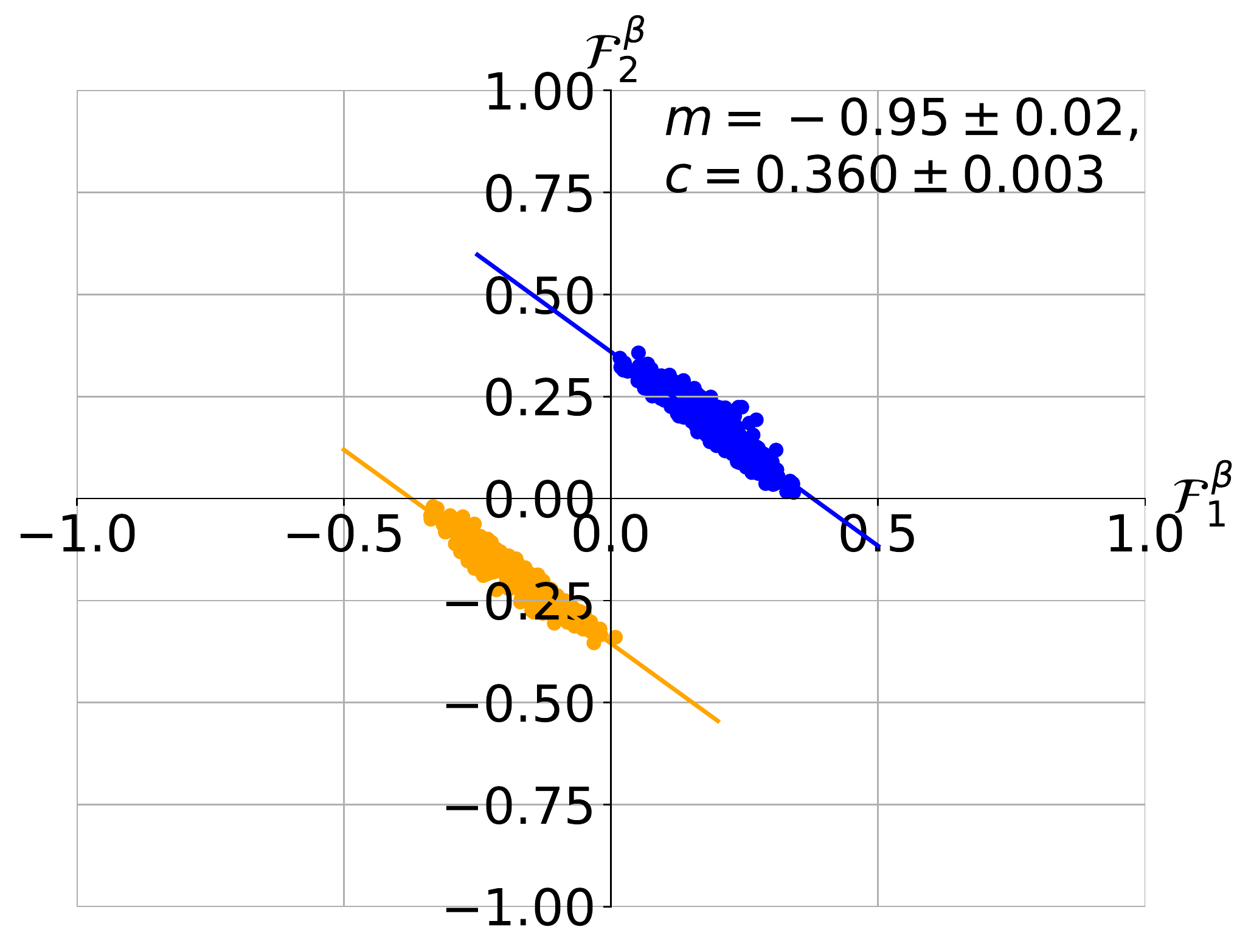}} %
\sidesubfloat[]{\includegraphics[width=0.4%
\columnwidth]{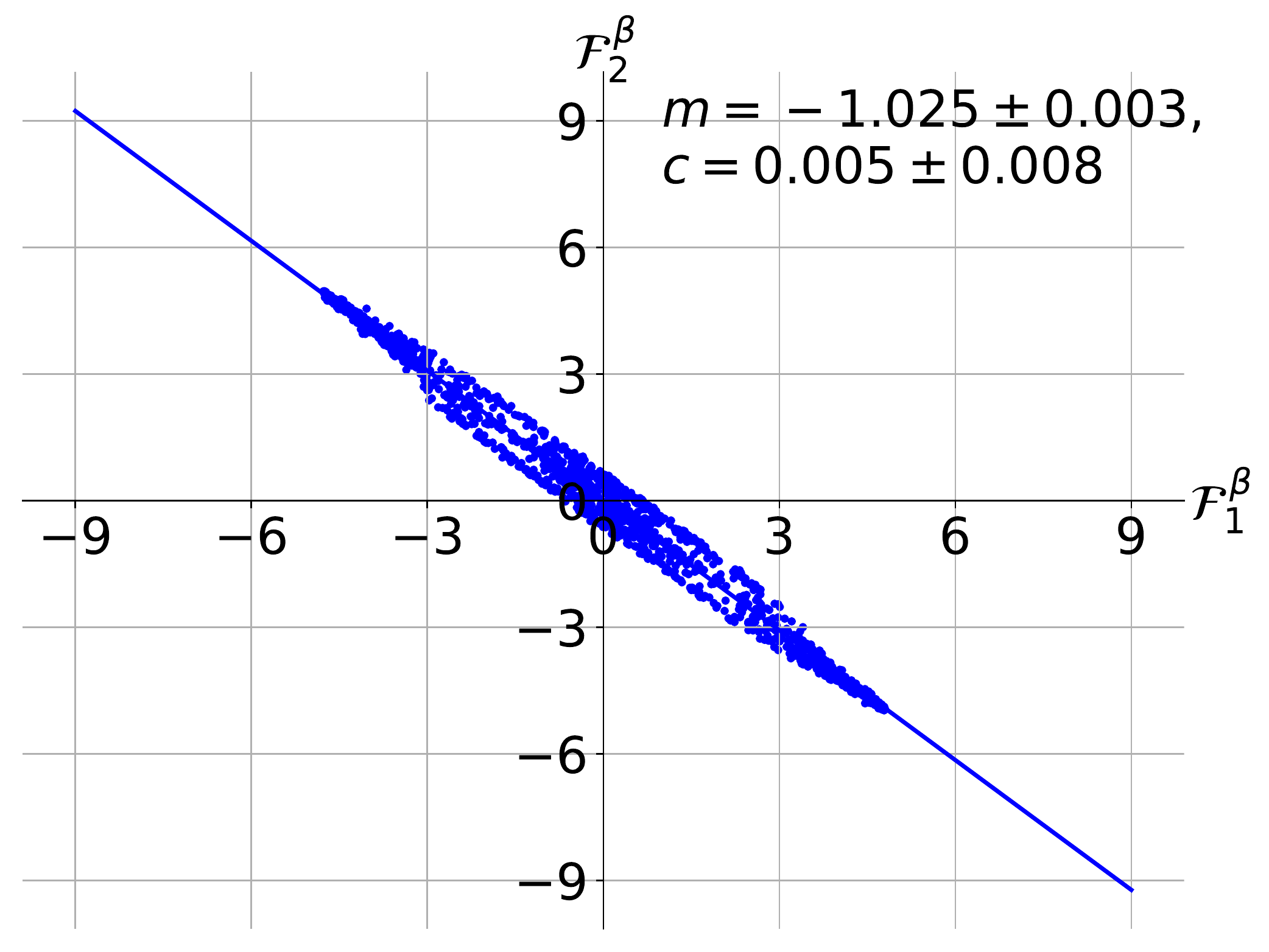}}
\caption{(a) Filter weights after performing convolution with a kernel of
width $2$ on the energy spectrum $\{E_{i}\}$ of 10 sites random field
Heisenberg chain. Data is shown for $\mathcal{T}_{\mathscr{L}}=1000$
training loops keeping stride $s=1$ and using sigmoid activation function.
Other training parameters are: batch size $N_{b}=100$, learning rate $%
\protect\eta =0.01$, and filter number $N_{f}=1$. (b) Filter weights for all
the filter channels obtained from normalized energy levels $\{E_{i}^{\prime
}\}$. Other parameters are: $N_{b}=100$, $s=1$, $\protect\eta =0.05$, $%
N_{f}=3$, and $\mathcal{T}_{\mathscr{L}}=500$. Here we use tanh as the
activation function.}
\label{Fig:C}
\end{figure}

In Fig.~\ref{Fig:C}(a) we show the resulting filter weights $\mathcal{F}%
^{\beta}$ from the $1000$ training process, in which we only use $N_f = 1$
filter. We find that the filter weights split into two branches in quadrants
I and III, respectively. We fit the filter weights by
\begin{equation}
\mathcal{F}_{1}^{\beta }= m \mathcal{F}_{2}^{\beta } \pm c,  \label{Eq:2}
\end{equation}%
where the plus sign corresponds to quadrant I and minus to quadrant III. The
best fit yields $m = -0.95 \pm 0.02$ and $c = 0.360 \pm 0.003$. We note that
$m \approx -1$ so the feature map, according to Eq.~(\ref{Eq:conv}), is
approximately
\begin{equation}
\mathrm{Z}_{i}^{\beta }=\mathcal{F}_{2}^{\beta }\Delta E_{i} \pm 0.36E_{i},
\label{Eq:4}
\end{equation}%
where $\Delta E_{i}=E_{i+1}-E_{i}$ is the nearest-neighbour level spacing.
Eq.~(\ref{Eq:4}) reveals two underlying features that the neural network
filters to distinguish the phases: $\Delta E_{i}$ and $E_i$. Because the
network identifies the phases with almost perfect accuracy regardless of the
value of $\mathcal{F}_{2}^{\beta}$, we conclude that the neural network is
not sensitive to level spacing, which is commonly used as a diagnostic for
phase transition in disordered and chaotic systems. The apparently
surprising result roots in the supervised learning scheme and the disorder
strength dependence of the bandwidth. As shown in Fig.~\ref{Fig:A}(a), the
bandwidth of the system grows monotonically with the disorder strength. In
the supervised learning bandwidth can be a feature that the CNN learns to
distinguish phases. For two neighbouring levels we can combine their
energies into their difference and mean, which are independent. Our
observation suggests that the energy difference is irrelevant, confirming
that their mean is the feature that is filtered by the convolution layer to
the subsequent fully connected layer, the output of which scales with
bandwidth.

This is a vivid example that a machine with a supervised scheme may not
always learn nontrivial knowledge. Bandwidth can be used to distinguish a
low-disorder system from a high disorder system, but it cannot be used to
detect the phase transition point without prior knowledge. The above example
is, therefore, a caution to the study of phase transition via deep neural
networks.

The failure to gain nontrivial features can be corrected, however, by
properly manipulating the input data. We can unbiasedly compare the spectra
at various disorder by normalizing the spectrum of a sample by its minimum
and maximum energies as
\begin{equation}
E_{i}^{^{\prime }}=2\frac{E_{i}-E_{\min }}{E_{\max }-E_{\min }}-1
\end{equation}%
where $E_{i}^{^{\prime }}$ is the normalized energy level. The normalized
spectrum $\left\{ E_{i}^{^{\prime }}\right\} $ always has a bandwidth of 2
and preserves the level statistics of the original spectrum. We then feed
the normalized spectrum $\left\{ E_{i}^{^{\prime }}\right\} $ to the CNN and
perform $500$ independent trainings $\mathcal{T}_{\mathscr{L}}$. We find
that once the bandwidth effects are removed by normalization the performance
of the same neural network drops from $100\%$ to around $70\%$, even though
we increase the number of channel to $N_f = 3$. Fig.~\ref{Fig:C}(b)
presents the results of the filter weights $\mathcal{F}^{\beta }$ in all
trainings, which fall roughly on a straight line. When we fit the data by
\begin{equation}
\mathcal{F}_{1}^{\beta }= m \mathcal{F}_{2}^{\beta } + c,  \label{Eq:2p}
\end{equation}%
we obtain $m = -1.025 \pm 0.003$ and $c = 0.005 \pm 0.008$. The
normalization suppresses the intercept to essentially zero, but preserves $m
\approx -1$. This means that the convolution layer now only passes the level
spacing information to the subsequent layer, consistent with our expectation
that level spacing can be used to detect the MBL phase and the MBL-thermal
transition.

In Fig.~\ref{Fig:C}(b) we use $N_f = 3$ convolutional filter channels. The
resulting weights for the three channels all prefer level-spacing information,
rather than bandwidth. We have also tested training with $N_f=1$ and $N_f=2$ channels.
We find, in general, more training loops (hence longer time) are needed to filter the bandwidth
information with less filters. For example, we need about 1000 training loops with $N_f=2$,
compared to 500 loops with $N_f = 3$ to obtain a similar trend line
as in Fig.~\ref{Fig:C}(b). We speculate that this is because the level spacing distribution
is a statistical ``order parameter'' in this system. Therefore, more
level-spacing information can be extracted with a larger $N_f$,
which allows an easier identification of its distribution.

We also note that the detail of the energy
normalization is not important. We also consider the normalization by $%
E_{i}^{^{\prime }}=\left( E_{i}-\overline{E}\right) /\sigma $ where $%
\overline{E}$ is the average energy of the spectrum and $\sigma $ is the
standard deviation from mean value. This normalization convention yields a
similar picture for the CNN parameters.

We can attribute the significant drop of the performance of the CNN after we
input the normalized spectra to sample-to-sample fluctuations in finite
systems. In the conventional level statistics study we identify the phase by
analyzing the ensemble averaged level spacing. In the CNN approach, however,
we ask which phase each sample belongs to. In a finite system, the
fluctuations in energy level spacing, therefore, prevent us from
unambiguously classifying individual samples. But over all samples, we still
achieve a 70\% performance, which is significantly higher than 50\% from
random guesses. The difference is sufficient for the machine to select the
level spacing as the feature map to explore during the convolution. The next
question is whether we can boost the performance by increasing the kernel
width.

\subsection{Filters whose kernel width is 3}

The CNN with kernel width 2 extracts the nearest-neighbour level spacing. To
extract more information we also study the neural network with filters of
kernel width 3, which captures both nearest-neighbour and
next-nearest-neighbour level spacings. Explicitly, a filter $\mathcal{F}%
^{\beta } = ( \mathcal{F}_{1}^{\beta }, \mathcal{F}_{2}^{\beta }, \mathcal{F}%
_{3}^{\beta })$ yields a feature map in the form
\begin{equation}
\mathrm{Z}_{i}^{\beta }=\mathcal{F}_{1}^{\beta }E_{i}+\mathcal{F}_{2}^{\beta
}E_{i+1}+\mathcal{F}_{3}^{\beta }E_{i+2}  \label{Eq:1}
\end{equation}%
Again, we input the energy spectra normalized by their minimum and maximum
energies. In this case, we obtain an $82\%$ performance in accuracy among
500 trainings, significantly higher than the 70\% with filters of kernel
width 2.

\begin{figure}[tbp]
\centering
\sidesubfloat[]{\includegraphics[width=0.4%
\columnwidth]{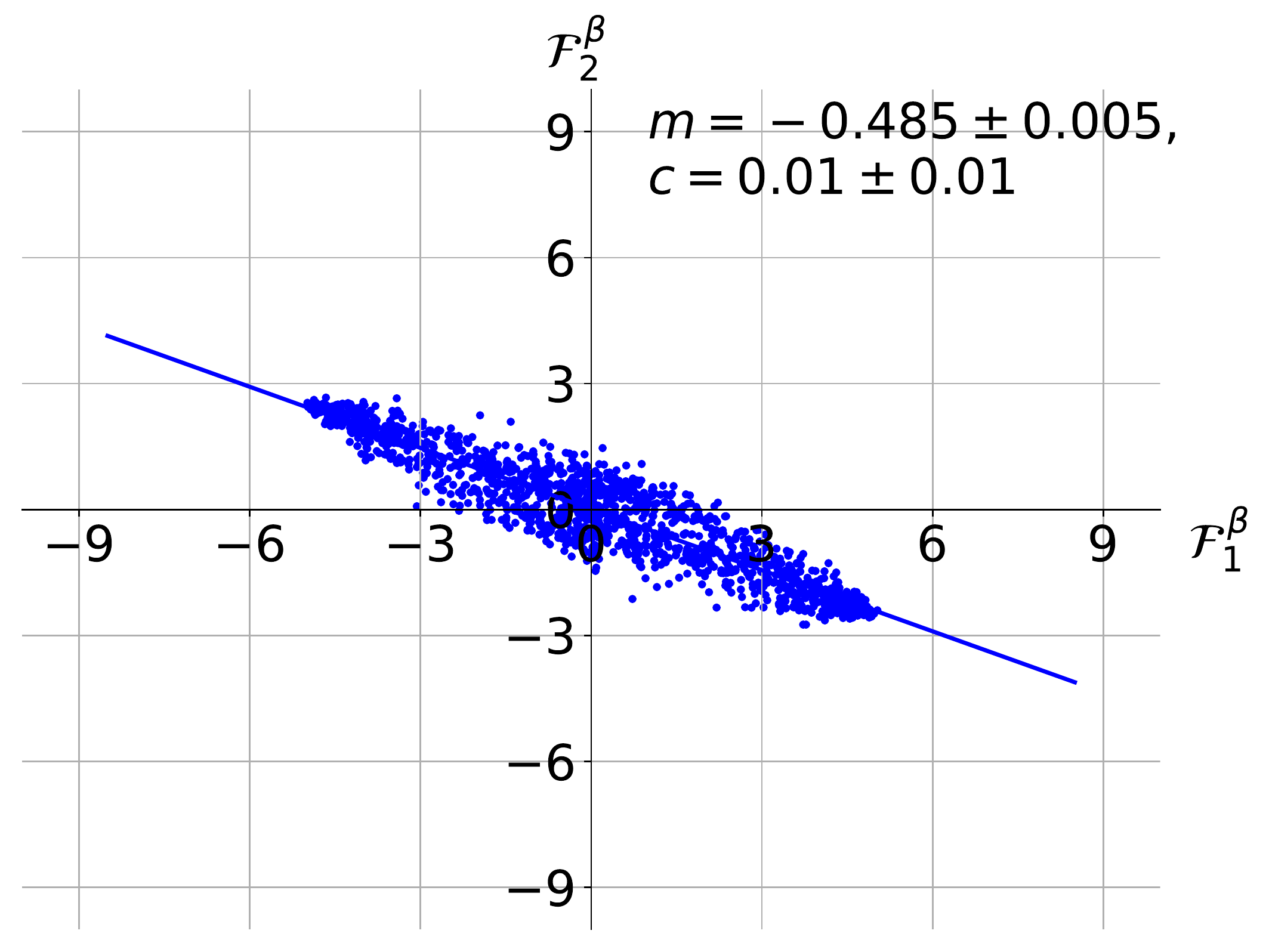}} %
\sidesubfloat[]{\includegraphics[width=0.4%
\columnwidth]{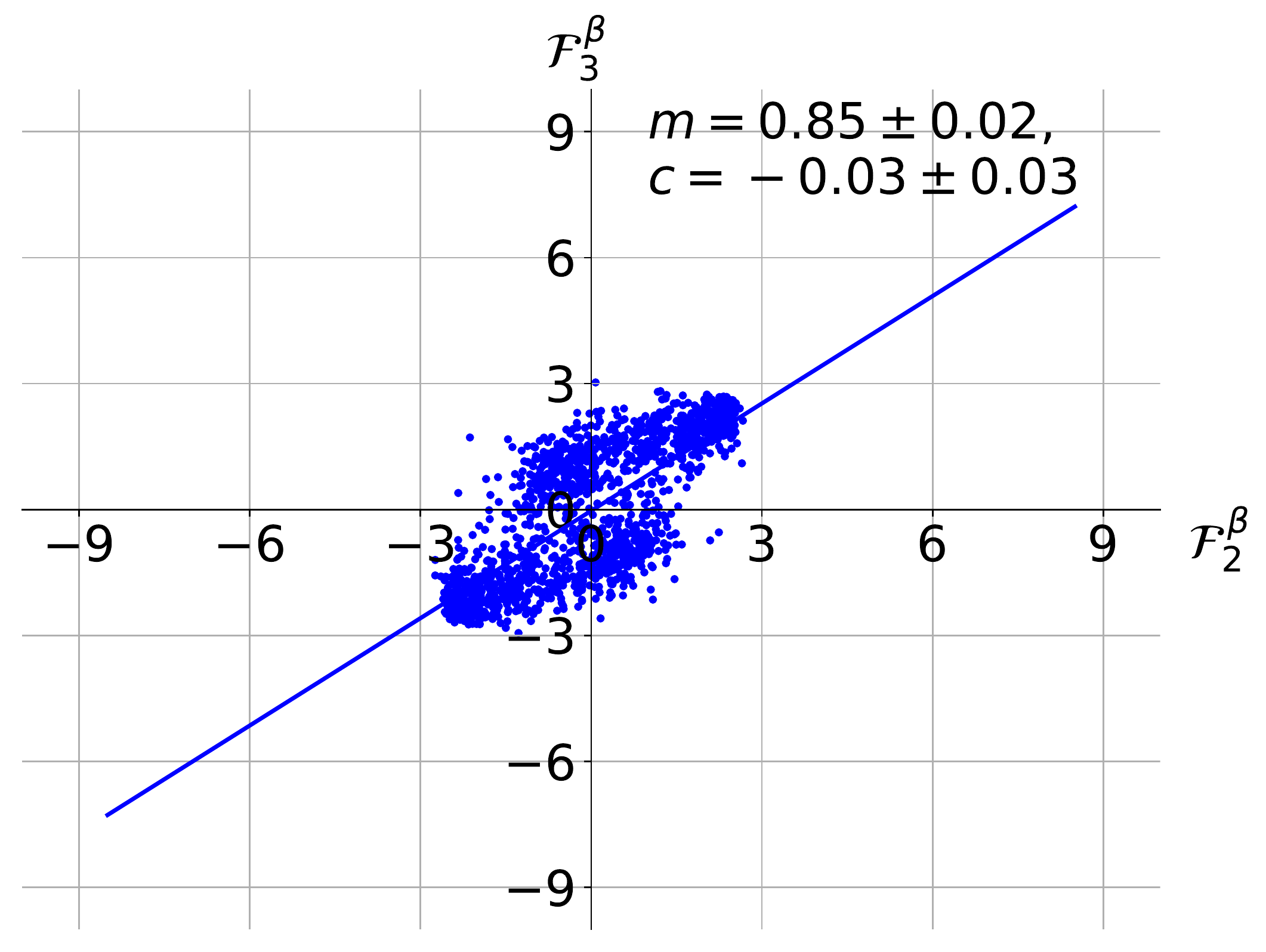}}
\caption{(a) Filter weights $\mathcal{F}_{2}^{\protect\beta }$ as a function
of $\mathcal{F}_{1}^{\protect\beta }$ obtained by using a kernel of size $%
3\times 1$ on normalized energy spectrum $\{E_{i}^{\prime }\}$, the stride
is kept to be $\mathrm{s}=1$. Other training parameters are: $N_{b}=100$, $%
\protect\eta =0.05$, $N_{f}=3$, and $\mathcal{T}_{\mathscr{L}}=500$. (b)
Filter weights $\mathcal{F}_{3}^{\protect\beta }$ as a function of $\mathcal{%
F}_{2}^{\protect\beta }$. Training parameters are the same as in (a).}
\label{Fig:D}
\end{figure}

We plot the resulting filter weights $\mathcal{F}_{2}^{\beta}$ against $%
\mathcal{F}_{1}^{\beta}$ in Fig.~\ref{Fig:D}(a) and $\mathcal{F}_{3}^{\beta}$
against $\mathcal{F}_{1}^{\beta}$ in Fig.~\ref{Fig:D}(b) and fit the results
by straight lines. We find that
\begin{equation}
\mathcal{F}_{2}^{\beta }= -(0.485 \pm 0.005) \mathcal{F}_{1}^{\beta } +
(0.01 \pm 0.01),
\end{equation}%
and
\begin{equation}
\mathcal{F}_{3}^{\beta }= (0.85 \pm 0.02) \mathcal{F}_{2}^{\beta } + (-0.03
\pm 0.03).
\end{equation}%
The results provide a strong motivation for us to approximate the weights by
\begin{equation}
\mathcal{F}_{2}^{\beta }=-\frac{\mathcal{F}_{1}^{\beta }}{2}\text{\ and\ }%
\mathcal{F}_{3}^{\beta } = \mathcal{F}_{2}^{\beta },
\end{equation}%
which lead to the feature map $\mathrm{Z}_{i}^{\beta }$ only in terms of one
filter element $\mathcal{F}_{1}^{\beta }$,
\begin{equation}
\mathrm{Z}_{i}^{\beta }\simeq \frac{\mathcal{F}_{1}^{\beta }}{2}%
(E_{i}-E_{i+1})+\frac{\mathcal{F}_{1}^{\beta }}{2}(E_{i}-E_{i+2})
\label{Eq:alpha3by1}
\end{equation}%
This feature map is an equal-weight linear combination of the
nearest-neighbour and next-nearest-neighbour level spacings.

The filter weights learned by the CNN suggests that the neural network, when
given the freedom to choose, tends to use both nearest-neighbour and
next-nearest-neighbour level spacings with roughly equal weight to identify
the energy spectrum of the two phases across the MBL phase transition. The
improved performance confirms that the combination is more effective than
using nearest-neighbour level statistics only. Therefore, we will turn to
the analysis of the next-nearest-neighbour level statistics in the next
subsection and try to understand the machine learning results with kernel
width 3.

We note that the training results scatter around the linear regression curve
with noticeable deviations. This results from the fact that the neural
network has numerous parameters or weights and, therefore, finding the
globally optimal, hence reproducible, parameters is almost impossible.
However, the goal of our study is not the precision of the parameters, but
the numerical trend that allows us to propose alternative level spacings,
which may outperform the conventional nearest-neighbour level spacings, at
least in finite systems.

\subsection{Next-nearest-neighbour level statistics}

Before we try to understand the machine learning results, we discuss the
distribution of the next-nearest-neighbour level spacings, which is denoted
by $\mathcal{P}_{2}(s)$. As for the nearest-neighbour level statistics, we
demand
\begin{equation}  \label{eq:normalization}
\int_0^{\infty} \mathcal{P}_{2}(s) ds = \int_0^{\infty} s \mathcal{P}_{2}(s)
ds = 1,
\end{equation}
which can be achieved by normalizing the level spacings by their mean.

In the MBL phase neighbouring eigenenergies likely correspond to two wave
functions localized in different regions. Therefore, we can write the
next-nearest-neighbour level spacing $E_{i+2}-E_{i}=\left(
E_{i+2}-E_{i+1}\right) +\left(E_{i+1}-E_{i}\right) $, as the sum of two
independent nearest-neighbour level spacings, whose distribution satisfies a
Poisson distribution $\mathcal{P} \left( s \right) =\exp \left( -s \right) $%
. Therefore, we have, for unnormalized level spacing $s^{\prime }$,
\begin{equation}
\tilde{\mathcal{P}}_{2}\left( s^{\prime }\right) \propto \int_{0}^{s^{\prime
}}\mathcal{P} \left( s^{\prime }-s_{1}\right) \mathcal{P} \left(
s_{1}\right) ds_{1}=s^{\prime} e^{ -s^{\prime }}\text{.}  \label{Eq:NNN}
\end{equation}%
Normalizing the distribution according to Eq.~(\ref{eq:normalization}), we
obtain
\begin{equation}
\mathcal{P}_{2}(s) = 4s\exp \left( {-2s}\right) ,
\end{equation}
which turns out to be a semi-Poisson distribution. Compared to the
nearest-neighbour level statistics, the most noticeable difference is now $%
\mathcal{P}_{2}\left( 0\right) =0$. This is not a manifestation of level
repulsion as in the thermal phase; rather, it simply states that three
consecutive levels do not coincide.

In the thermal phase neighbouring levels are correlated. In random matrix
theory the joint probability density function for eigenvalues is \cite%
{Haake2001}
\begin{equation}
P(\{E_{i}\})\propto \prod_{i<j}|E_{i}-E_{j}|^{\nu }\exp (-A\sum_{i}E_{i}^{2})%
\text{,}
\end{equation}%
where $\nu =1$ in the GOE. We show, in \ref{app:nnn}, that the distribution
leads to the distribution of next-nearest-neighbour level spacings
\begin{equation}
\mathcal{P}_{2}(s)=\frac{2^{18}}{3^{6}\pi ^{3}}s^{4}\exp \left( -\frac{64}{%
9\pi }s^{2}\right) \text{,}  \label{Eq:NNN2}
\end{equation}%
which is, interestingly, identical to the distribution of nearest-neighbour
level spacings in a GSE. In contrast, if we neglect the correlations between
neighbouring level spacings and adopt similar derivations as Eq.~(\ref%
{Eq:NNN}), the distribution of next-nearest-neighbour level spacings is
\begin{equation}
\mathcal{P}_{2}^{^{\prime }}\left( s\right) =\pi se^{-\pi s^{2}}+\frac{\pi
\left( \pi s^{2}-1\right) }{\sqrt{2}}e^{-\pi s^{2}/2}\text{Erf}\left( \sqrt{%
\frac{\pi }{2}}s\right) \text{,}  \label{Eq:dist}
\end{equation}%
where Erf stands for the error function Erf$\left( z\right) =\frac{2}{\sqrt{%
\pi }}\int_{0}^{z}e^{-t^{2}}dt$.

\begin{figure}[tbp]
\centering
\includegraphics[width=0.45\columnwidth]{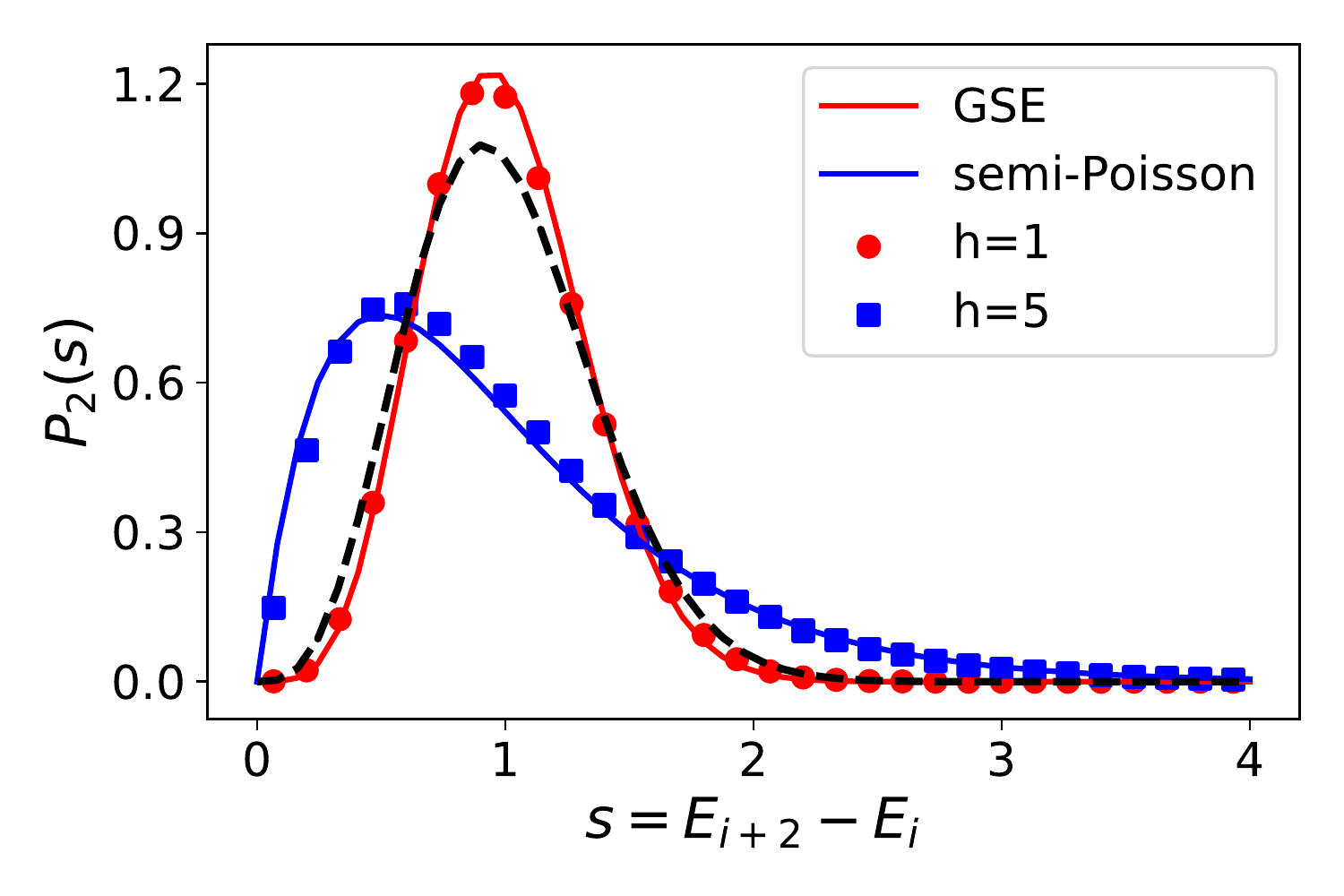}
\caption{Evolution of next-nearest level spacing distribution from GOE to
GSE at $h=1$ and from Poisson to semi-Poisson at $h=5$. The black dashed
line corresponds to the distribution in Eq. (\protect\ref{Eq:dist}).}
\label{Fig:E}
\end{figure}

In Fig.~\ref{Fig:E} we plot the distribution of next-nearest-neighbour level
spacings of the random spin chains. We confirm that the distribution for $h
= 1$ agrees well with the GSE distribution, while that for $h = 5$ follows
the semi-Poisson distribution. For comparison, we also plot Eq.~(\ref%
{Eq:dist}) by a black dashed line, which clearly deviates from the data at $%
h = 1$, indicating that the correlation in level spacings is non-negligible.

Comparing Fig.~\ref{Fig:E} to Fig.~\ref{Fig:A}(b), we conclude that the
distribution of next-nearest-neighbour level spacings can also be used to
distinguish the energy spectra in the thermal and MBL phases. Because in the
MBL phase the correlation between eigenstates decays exponentially with
their distance, it is more advantageous to consider the distribution of
next-nearest-neighbour level spacings than that of nearest-neighbour level
spacings in finite systems, which, at large disorder, deviates from the
Poisson distribution at small level spacings, as shown in Fig.~\ref{Fig:A}%
(b).

However, the comparison is for the ensemble average. On the other hand, ML
tends to ask whether an individual sample belongs to the thermal or the MBL
phase, and the phase boundary can be determined by the percentage
recognition of the two phases. In this case, our CNN study in the previous
subsection reveals roughly equal weights on the two level spacings, which
improves the recognition accuracy significantly. We speculate that while the
residual level repulsion due to the finite-size effect in the MBL phase
favours the next-nearest-neighbour level statistics, the fluctuations in the
level spacings tend to blur the difference in the similar peak structures in
the two phases. So what we observe is likely a compromise of the two aspects
of the finite-size effect. Finally, it's worth emphasizing the filter size
can be further increased to incorporate level spacings on longer ranges,
and the higher-order level spacings are also efficient for distinguishing
phases. Meanwhile, the results from filter size equals two and three are
sufficient to express the main point of current work.

\section{Conclusion}

We deploy a CNN to study the thermal-MBL transition in a one-dimensional
random spin system, using the raw energy spectrum as the training data. Our
aim is to reveal the key feature that the neural network extracts to
classify the phases. The simplest CNN that contains $2\times 1$ filters can
capture the nearest-neighbour level spacings, which can be used to
distinguish the thermal phase from the MBL phase. However, the accuracy to
identify individual samples is limited by finite-size effect. By using $%
3\times1$ filters, the CNN is able to capture next-nearest-neighbouring
level spacings. We compare the next-nearest-neighbor level statistics for
the thermal and MBL phases to analytical solutions and show that it can also
be used to distinguish the two phases. As a result, the CNN improves the
test accuracy by $12\%$ by enlarging filters.

Compared to earlier studies, the present approach has the following
advantages. First, we use raw energy spectra (normalized to get rid of the
trivial band width information in the supervised learning) as the training
data. Compared to ES that was used in Ref.~\cite{Nieuwenburg17}, our training data is considered to
be of \textquotedblleft lower level\textquotedblright, which requires less
prior knowledge. Second, although the fully-connected NN has
been used in learning the thermal-MBL transition\cite{Rao182}, the large number of
parameters makes it difficult to extract quantities of physical meaning that
machine utilizes to distinguish phases. The filters in our CNN architecture
allows a clear interpretation of what a machine learns.
In contrast, the training with normalized energy spectrum
turns out to be much more difficult and unstable for the fully-connected NN.
For a comparison, the learning rate in Ref.~\cite{Rao182} is the order of $10^{-4}$,
which should be reduced to the order of $10^{-8}$ when training data is the
normalized energy spectrum. This demonstrates that the NN without a feature extractor
is less efficient in learning non-trivial physics, which
motivated us to employ a CNN.
Finally, we are able to show that machine learning can \textquotedblleft discover\textquotedblright
less known physical quantity, in this case higher-order level spacings for
both the thermal and MBL phases.
Therefore, the present work provides a vivid example of how one
may use neural networks to develop and to improve
methods from low-level data in disordered systems.

In general, our approach can be applied to study dynamical phase transition
in any model that has energy or entanglement spectrum. For example, in
quantum system with periodic driving where a quasi-energy spectrum replaces
the conventional energy spectrum, we believe the CNN can likewise capture
the dynamical signal of phase transition through the filters. In addition,
by selecting different parts of the energy spectrum as training data, the CNN
can also be used to locate the many-body mobility edge.

We note that our discussion only relies on random matrix theory, rather than
the specific Hamiltonian. We expect the results can be applied to other
disordered and chaotic systems in both ML and conventional studies.

\section*{Acknowledgement}

We acknowledge the support by the National Natural Science Foundation of
China through Grant No. 11904069, No. 11847005 and No. 11674282 and the
Strategic Priority Research Program of Chinese Academy of Sciences through
Grant No. XDB28000000.

\appendix

\section{Derivation of Eq.~(\protect\ref{Eq:NNN2})}

\label{app:nnn}

In this appendix we give an analytical derivation for the next-nearest level
spacings in thermal phase. We start with the standard (unnormalized) energy
level probability density for Gaussian ensembles~\cite{Haake2001},%
\begin{equation}
P\left( \left\{ E_{i}\right\} \right) \propto \prod_{i<j}\left\vert
E_{i}-E_{j}\right\vert ^{\nu }e^{-A\sum_{i}E_{i}^{2}}
\end{equation}%
where $\nu =1$, 2, 4 for GOE, GUE, and GSE, respectively. When dealing with
nearest-neighbour level spacing, it is sufficient to consider the $2\times 2$
matrix case~\cite{Haake2001}. Likewise, to study the next-nearest level
spacing, we can consider $3\times 3$ matrix. Introduce $\mathcal{P}%
_{2}\left( s\right) =P\left( \left\vert E_{3}-E_{1}\right\vert =s\right) $,
we have%
\begin{equation}
\mathcal{P}_{2}\left( s\right) \propto \int_{-\infty }^{\infty
}\prod_{i<j}\left\vert E_{i}-E_{j}\right\vert ^{\nu }\delta \left(
s-\left\vert E_{1}-E_{3}\right\vert \right)
e^{-A\sum_{i}E_{i}^{2}}dE_{1}dE_{2}dE_{3}\text{.}
\end{equation}%
Now we switch variables to $x_{1}=E_{1}-E_{2}$, $x_{2}=E_{2}-E_{3}$, $%
x_{3}=\sum_{i=1}^{3}E_{i}$, then%
\begin{eqnarray}
\mathcal{P}_{2}\left( s\right)  &\propto &\int_{-\infty }^{\infty
}\left\vert x_{1}\right\vert ^{\nu }\left\vert x_{2}\right\vert ^{\nu
}\left\vert x_{1}+x_{2}\right\vert ^{\nu }\delta \left( s-\left\vert
x_{1}+x_{2}\right\vert \right) e^{-\frac{A}{2}\left[
x_{1}^{2}+x_{2}^{2}+x_{3}^{2}+\left( x_{1}+x_{2}\right) ^{2}\right] }
\nonumber \\
&&\times \frac{\partial \left( E_{1},E_{2},E_{3}\right) }{\partial \left(
x_{1},x_{2},x_{3}\right) }dx_{1}dx_{2}dx_{3}\text{.}
\end{eqnarray}%
In this integral the Jacobian $\frac{\partial \left(
E_{1},E_{2},E_{3}\right) }{\partial \left( x_{1},x_{2},x_{3}\right) }$ and
the integral for $x_{3}$ are all constants that can be absorbed into the
normalization factor. By introducing the polar coordinates $x_{1}=r\cos
\theta $, $x_{2}=r\sin \theta $, we can write $\mathcal{P}_{2}\left(
s\right) $ as%
\begin{eqnarray}
\mathcal{P}_{2}\left( s\right)  &\propto &\int_{0}^{\infty }\int_{0}^{2\pi
}r^{3\nu }\left\vert \cos \theta \right\vert ^{\nu }\left\vert \sin \theta
\right\vert ^{\nu }\left\vert \cos \theta +\sin \theta \right\vert ^{\nu }
\nonumber \\
&&\times \delta \left( s-r\left\vert \cos \theta +\sin \theta \right\vert
\right) e^{-\frac{A}{2}r^{2}\left( 2+\sin 2\theta \right) }rdrd\theta
\nonumber \\
&=&\int_{0}^{2\pi }\left( \frac{s}{\left\vert \cos \theta +\sin \theta
\right\vert }\right) ^{3\nu +1}e^{-\frac{A\left( 2+\sin 2\theta \right) }{%
2\left\vert \cos \theta +\sin \theta \right\vert ^{2}}s^{2}}  \nonumber \\
&&\times \left\vert \cos \theta \right\vert ^{\nu }\left\vert \sin \theta
\right\vert ^{\nu }\left\vert \cos \theta +\sin \theta \right\vert ^{\nu
}d\theta \text{.}
\end{eqnarray}%
Although the integral for $\theta $ is difficult to solve, it only
contributes to the normalization factor and does not influence the scaling
behavior of $s$. Therefore, we can simplify $\mathcal{P}_{2}\left( s\right) $
to
\begin{equation}
\mathcal{P}_{2}\left( s\right) =C\left( \nu \right) s^{3\nu +1}e^{-A\left(
\nu \right) s^{2}}\text{.}
\end{equation}%
For GOE, we have $\nu =1$. Finally, by imposing the normalization condition%
\begin{equation}
\int_{0}^{\infty }\mathcal{P}_{2}\left( s\right) ds=1\text{, }%
\int_{0}^{\infty }s\mathcal{P}_{2}\left( s\right) =1
\end{equation}%
we can determine the coefficients $C\left( \nu \right) $ and $A\left( \nu
\right) $ and obtain the GSE distribution as in Eq.~(\ref{Eq:NNN2}). The
higher-order level spacing distributions can be derived in a similar manner.

\end{document}